\documentclass[epj]{svjour}
\usepackage{graphicx}

\begin{document}

\title{Neutron Transfer Dynamics and Doorway to Fusion\\
       in Time-Dependent Hartree-Fock Theory}
%\author{First author\inst{1} \and Second author\inst{2}% multiple authors.
\author{A.S. Umar\inst{1}\thanks{\emph{e-mail:} umar@compsci.cas.vanderbilt.edu}, V.E. Oberacker\inst{1},
\and J.A. Maruhn\inst{2}}
% When running author/title are too long use \authorrunning and \titlerunning to give shorter versions
\authorrunning{Umar, Oberacker, Maruhn}
\titlerunning{Transfer Dynamics and Doorway to Fusion in TDHF}

\institute{Department of Physics and Astronomy, Vanderbilt University, Nashville, Tennessee 37235, USA
\and Institut f\"ur Theoretische Physik, Goethe-Universit\"at, 60438 Frankfurt am Main, Germany}
%\institute{Insert the first address here \and the second here}

\date{Received: \today / Revised version: date}

\abstract{
We analyze the details of mass exchange in the vicinity of the Coulomb barrier for heavy-ion collisions
involving neutron-rich nuclei using the time-dependent Hartree-Fock (TDHF) theory. We discuss the time-dependence
of transfer and show that the potential barriers seen by individual single-particle
states can be considerably different than the effective barrier for the two interacting
nuclei having a single center-of-mass.
For this reason we observe a substantial transfer probability even at energies below the effective barrier.
\PACS{{21.60.Jz}{Nuclear Density Functional Theory and extensions} \and {24.10.Cn}{Many-body theory}} % end of PACS codes
} %end of abstract

\maketitle

%------------------------------------------------------------------------------

\section{Introduction}

Heavy-ion fusion reactions are a sensitive probe of the size, shape, and structure
of atomic nuclei as well as the collision dynamics.
With the increasing availability of radioactive ion-beams the study of
fusion reactions of neutron-rich nuclei are now possible~\cite{Li03,Li05,Ji04}.
Other experimental frontiers are the synthesis of superheavy nuclei in cold and hot fusion
reactions~\cite{Ho02,Og04,Gi03,Mo04,II05}, and weakly bound light systems~\cite{YZ06,PF04,KG98,TS00}.
Microscopic descriptions of nuclear fusion may provide us with a better understanding of the
interplay between the strong, Coulomb, and the weak interactions as well as the enhanced
static and dynamic correlations present in these many-body systems.

Recently, two aspects of the collision dynamics leading to fusion that involve
pre-compound neutrons have been of interest.
Over the last decade a number of fusion studies have reported
that the average number of neutrons evaporated by the compound nucleus is
considerably less than what is predicted by statistical fusion evaporation
calculations~\cite{WB06}.
This phenomenon is quite possibly linked to the excitation of the pre-compound collective
dipole mode, which is likely to occur when ions have significantly different $N/Z$
ratio, and is a reflection of dynamical charge equilibration. This was studied
in the context of TDHF in Refs.~\cite{SC01,SC07,US85} and we have recently observed
this phenomenon in the $^{64}$Ni+$^{132}$Sn system~\cite{UO07a}.
Similarly, considerable attention has been given to the influence of neutron
transfer on fusion cross-sections. Studies suggest that the transfer of neutrons with
positive $Q$ value strongly enhances the fusion cross-section in comparison to systems
having negative $Q$ value~\cite{Li07,ZSW07,DLW83}. This may explain the fact that
lowering of the potential barrier for neutron-rich systems does not always lead to
higher fusion cross-sections.
In Ref.~\cite{ZSW07} near-barrier fusion of neutron-rich nuclei was studied within a
channel coupling model for intermediate neutron rearrangement using
a semi-empirical time-dependent three-body Schr\"odinger equation. Studies showed
that for the $^{40}$Ca+$^{96}$Zr system neutrons were transferred in the early
stages of the collision from the $2d_{5/2}$ state of $^{96}$Zr to the unoccupied
levels of the $^{40}$Ca nucleus.

It is generally acknowledged that the TDHF method provides a
useful foundation for a fully microscopic many-body theory of low-energy heavy-ion reactions~\cite{Ne82}.
Historically, fusion in TDHF has been viewed as a final product of two
colliding heavy-ions, and the dynamical details influencing the formation of the
compound system have not been carefully dissected in terms of the pre-compound properties.
Due to the availability of much richer fusion data and considerable advances in TDHF codes
that make no symmetry assumptions and use better effective interactions, it may now be possible
to examine these effects more carefully.
In TDHF complete fusion proceeds by converting the entire relative kinetic
energy in the entrance channel into internal excitations of a single well-defined compound nucleus.
The dissipation of the relative kinetic energy into internal excitations is
due to the collisions of the nucleons with the ``walls'' of the
self-consistent mean-field potential. TDHF studies demonstrate that the
randomization of the single-particle motion occurs through repeated exchange of
nucleons from one nucleus into the other. Consequently, the equilibration of
excitations is very slow, and it is sensitive to the details of the
evolution of the shape of the composite system.
This is in contrast to
most classical pictures of nuclear fusion, which generally assume near
instantaneous, isotropic equilibration.
The relaxation of the final compound system is a long-time process occurring on
a time scale on the order of a few thousand {\it fm/c}. In contrast, the
pre-compound stage corresponds to a time scale of a few hundred {\it fm/c}.

In this manuscript we focus on the analysis of transfer during the
early stages of the collision. In particular, we confirm the
findings of Ref.~\cite{ZSW07}. We also show that in TDHF different
single-particle states seem to see different potential barriers in
comparison to the generic ion-ion barrier. This influences the overall
dynamics leading to fusion and consequently the effective potential
barrier.

\section{Transfer in TDHF}

The TDHF calculations have been carried out using our new three-dimensional unrestricted
TDHF code~\cite{UO06}.
For the effective interaction we have used the Skyrme SLy4 force~\cite{CB98},
including all of the time-odd terms.
Static Hartree-Fock calculations for all the nuclei studied here produce
spherically symmetric systems.
The chosen mesh spacing was $1$~fm in all three directions, which yield a binding
energy accuracy of about $50$~keV in comparison to a spherical Hartree-Fock code.
For these calculations we have in addition required that the fluctuations in energy
be as low as $10^{-4}$-$10^{-5}$, the corresponding accuracy in binding energy is about $10^{-12}$.
This ensures that the tails of the wavefunctions are well converged in the numerical
box. The box size used was $60$~fm in the direction of the collision axis and $30$~fm
in the other two directions. The initial nuclear separations were $25$~fm.
\begin{figure}[!htb]
\includegraphics*[scale=0.42]{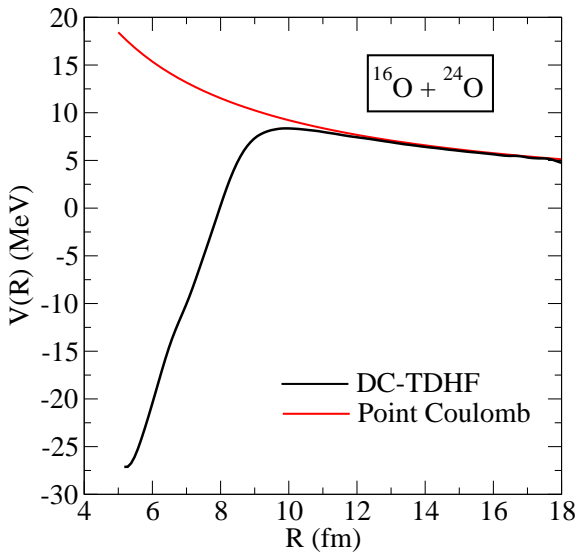}
\caption{\label{fig:vrOO} Potential barrier, $V(R)$, for the $^{16}$O+$^{24}$O system
obtained from density constrained TDHF calculations (black curve). Also shown is the point Coulomb potential (red curve).}
\end{figure}

\subsection{$^{16}$O+$^{24}$O system}

As an example of a collision involving one neutron rich nucleus we studied the $^{16}$O+$^{24}$O system.
In order to determine the potential barrier for the system we have used the DC-TDHF method as described in
Ref.~\cite{UO06b}. In this approach the TDHF time-evolution takes place with no restrictions.
At certain times during the evolution the instantaneous density is used to
perform a static Hartree-Fock minimization while holding the neutron and proton densities constrained
to be the corresponding instantaneous TDHF densities. Some of the effects naturally included in the
DC-TDHF calculations are: neck formation, particle exchange, internal excitations, and deformation
effects to all order.
The heavy-ion potential was obtained by initializing the system at $E_{\mathrm{c.m.}}=9.5$~MeV,
which is slightly above the barrier shown in Fig.~\ref{fig:vrOO}. The peak of the barrier is about
$8.4$~MeV at a nuclear separation of $9.9$~fm. This is lower than the
barrier of the $^{16}$O+$^{16}$O system,
which has a height of about $10$~MeV. 
Here and in the following, the heavy-ion interaction potential
has been calculated with a constant mass parameter corresponding to the reduced mass of the ions. This
is a good approximation as long as one is only interested in the value of the potential barrier height
(as is the case here). For the calculation of sub-barrier fusion cross sections, however, it is essential
that coordinate-dependent mass parameters be utilized~\cite{UO07a} because the cross sections depend
sensitively on the shape of the potential in the interior region.
\begin{figure}[!htb]
\includegraphics*[scale=0.42]{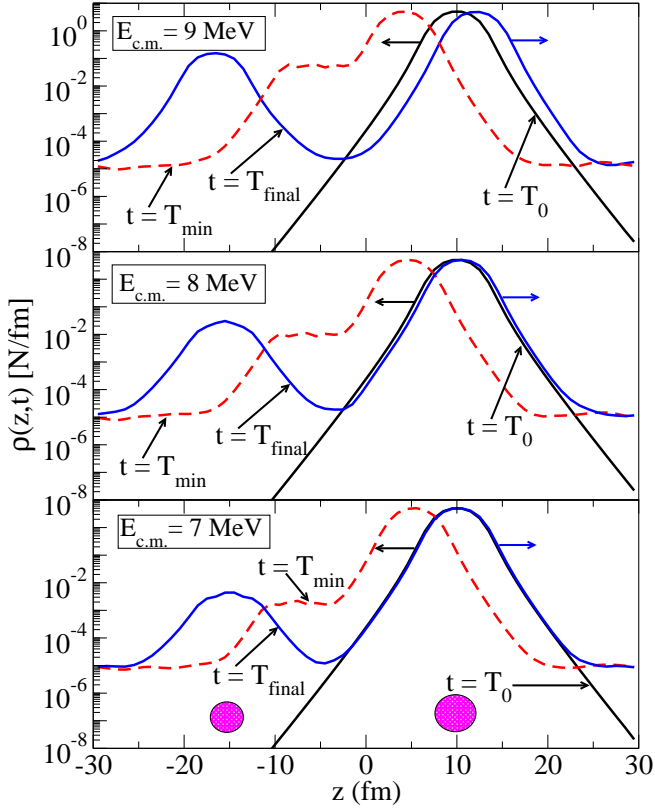}
\caption{\label{fig:rhoz} Partially integrated neutron densities calculated from Eq.(\protect\ref{eq:rhoz})
for the $^{24}$O nucleus
plotted on a logarithmic scale versus the collision axis coordinate $z$ for the $^{16}$O+$^{24}$O system
at three energies, $E_{\mathrm{c.m.}}=7,8$, and $9$~MeV. The black-solid curves correspond to the initial partial
density, the red-dashed curves are the same quantity at the distance of closest approach, and the blue-solid
curves are partial densities long after the recoil. Filled spheres near the bottom axis approximately show the
initial and final location of the two nuclei.}
\end{figure}

In order to examine the center-of-mass energy dependence of mass exchange below and above the
barrier we have initiated TDHF collisions at energies
$E_{\mathrm{c.m.}}=6$, 7, 8, 9, and 9.5~MeV.
Interestingly, the head-on (zero impact parameter) TDHF collisions for
the lowest four energies behave as a typical sub-barrier collision: the two ions approach a minimum distance
with no visible overlap, then recoil and move away from each other. This is also true for $E_{\mathrm{c.m.}}=9$~MeV
despite the fact that this energy lies above the ion-ion barrier.
This suggests that while we can talk about an {\it effective} ion-ion barrier the individual single-particle
states may see a barrier somewhat different than the effective one. This is in agreement with the findings
of Ref.~\cite{ZSW07}, and we shall come back to this point again later in the manuscript.
Even though we are dealing with sub-barrier energies, we observe mass exchange (mainly neutron)
from $^{24}$O to $^{16}$O for all these energies. Similarly, we observe a small dissipation
of the relative kinetic energy, ranging from $0.05$~MeV for $E_{\mathrm{c.m.}}=6$~MeV to $0.40$~MeV for
$E_{\mathrm{c.m.}}=9$~MeV. For comparison, the difference in total energy of the system before and after
the collision is on the order of $0.02$~MeV, i.e., the numerical error in total energy conservation is less than
the dissipated energy, even at very low energies. At $E_{\mathrm{c.m.}}=9.5$~MeV the system fuses.
\begin{figure}[!htb]
\includegraphics*[scale=0.45]{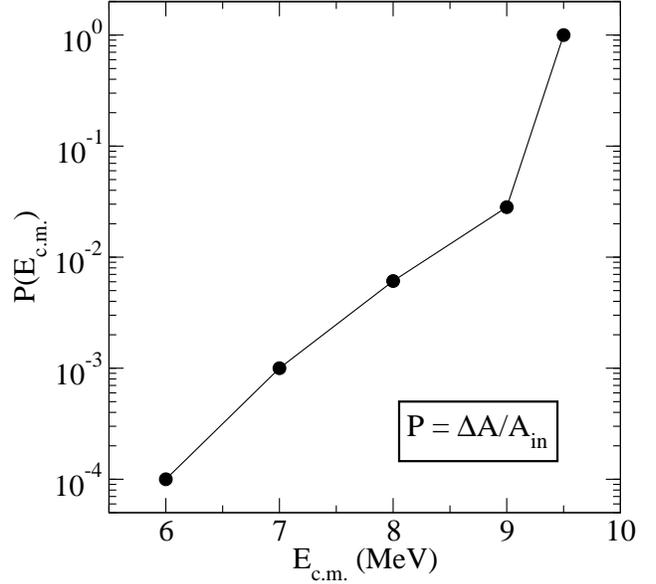}
\caption{\label{fig:pa} Mass transfer probability defined in Eq.~(\protect\ref{eq:pa}) for the transfer in the
$^{16}$O+$^{24}$O system.}
\end{figure}

To gain further insight into the mass exchange we have looked at the change in nuclear density
along the collision axis ($z$-axis). We define
\begin{equation}
\rho(z,t)=\int dx\int dy \; \rho(x,y,z,t)\;,
\label{eq:rhoz}
\end{equation}
where $\rho(x,y,z,t)$ is the instantaneous TDHF density.
In Fig.~\ref{fig:rhoz} we plot $\rho(z,t)$ for three energies,
$E_{\mathrm{c.m.}}=7$, 8, and $9$~MeV,
on a logarithmic scale versus the collision axis coordinate $z$. We emphasize that the plotted density 
contains only states that correspond to the $^{24}$O nucleus.
The black-solid curves correspond to the initial partial
density, the red-dashed curves are the same quantity at the distant of closest approach, and the blue-solid
curves are partial densities long after the recoil. Filled spheres near the bottom axis denote the approximate
initial and final location of the two nuclei. The distance of closest
approach for energies $E_{\mathrm{c.m.}}=$7, 8, and 9~MeV are 13.2~fm, 11.6~fm, and 10.1~fm, respectively.
Two features are worth noting: first, we see a buildup of the density on the $^{16}$O side of
the reaction plane, and secondly, we note that the buildup gets substantially larger with increasing
energy.
We also observe that the recoiled density profile remains very close to the initial profile except
in the tail region. The corresponding mass transfer $\Delta A$ at sub-barrier energies
can be found by integrating $\rho(z,t)$ over the $z$ values on the left side of the minimum. At energies
$E_{\mathrm{c.m.}}=7$, 8, and 9~MeV we find mass transfer values of 0.024, 0.1464, and 0.6768, respectively.
The mass transfer at $E_{\mathrm{c.m.}}=6$~MeV is 0.0022.
We emphasize that unitarity (or total mass number) was conserved
with an accuracy of about one part in $10^5$.
In Fig.~\ref{fig:pa} we plot the energy dependence of the transfer probability defined as
\begin{equation}
P(E_{\mathrm{c.m.}})=\frac{\Delta A(E_{\mathrm{c.m.}})}{A_{in}}\;,
\label{eq:pa}
\end{equation}
where $\Delta A$ is the mass exchange value mentioned above and $A_{in}$ is the mass of the incoming
nucleus, in this case $24$.
We observe that for sub-barrier energies the curve is essentially linear and jumps to one
when fusion occurs.
\begin{figure}[!htb]
\includegraphics*[scale=0.47]{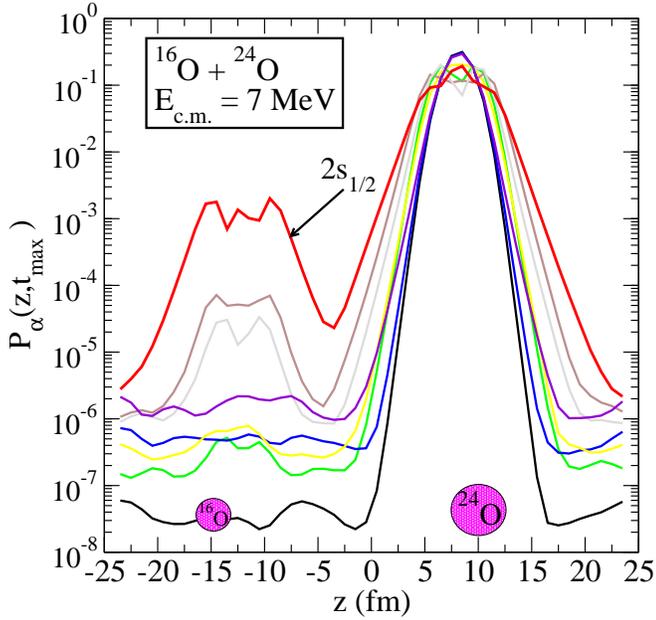}
\caption{\label{fig:palpha} Neutron single-particle probability densities given by Eq.~(\protect\ref{eq:palpha})
for the $^{24}$O nucleus at $E_{\mathrm{c.m.}}=7$~MeV at $t=T_{final}$. Filled spheres near the bottom axis approximately show the
final location of the two nuclei.}
\end{figure}

In order to identify which of the single-particle states are predominantly responsible for transfer at
sub-barrier energies we define the quantity
\begin{equation}
P_{\alpha}(z,t)=\int dx \int dy \; |\psi_{\alpha}(x,y,z,t)|^2\;,
\label{eq:palpha}
\end{equation}
where the $\psi_{\alpha}$'s are single-particle states. In Fig.~\ref{fig:palpha} we plot this quantity
on a logarithmic scale for the neutron states of the $^{24}$O nucleus at $E_{\mathrm{c.m.}}=7$~MeV, long after the recoil,
at which time the ion-ion separation is about $R=25$~fm.
Although our single-particle wave functions are calculated on a 3-D Cartesian grid and thus do not
carry the same good quantum numbers as the spherical representation, for a well converged spherical
nucleus it is possible to calculate the expectation values of orbital angular momentum, spin, and parity to
identify these states. As expected, the contribution to the sub-barrier mass transfer is primarily coming from the
$2s_{1/2}$ neutron state of the $^{24}$O nucleus. The contribution from the $d_{5/2}$ and $d_{3/2}$ states is
an order of magnitude lower than that of the $2s_{1/2}$ state. This again suggests that the barrier seen by the
$2s_{1/2}$ state is different than the effective barrier for the entire nucleus.
\begin{figure}[!htb]
\includegraphics*[scale=0.42]{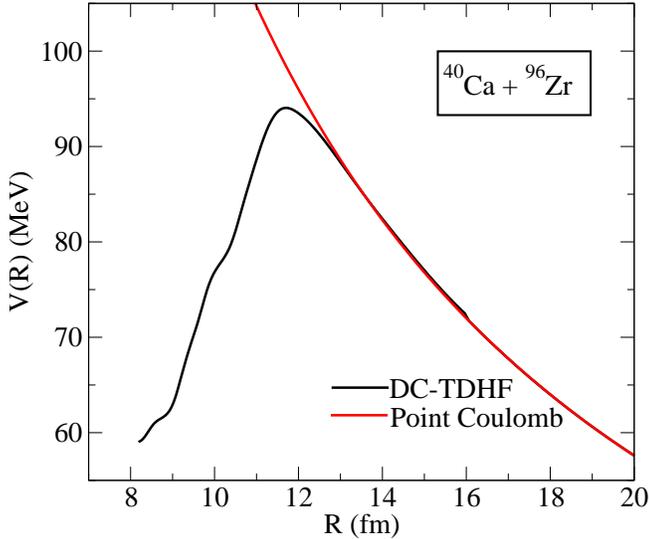}
\caption{\label{fig:vrCaZr} Potential barrier, $V(R)$, for the $^{40}$Ca+$^{96}$Zr system
obtained from density constrained TDHF calculations (black curve). Also shown is the point Coulomb potential (red curve).}
\end{figure}

\subsection{$^{40}$C\lowercase{a}+$^{96}$Z\lowercase{r} system}

\begin{figure}[!hbt]
\begin{center}
\includegraphics*[scale=0.55]{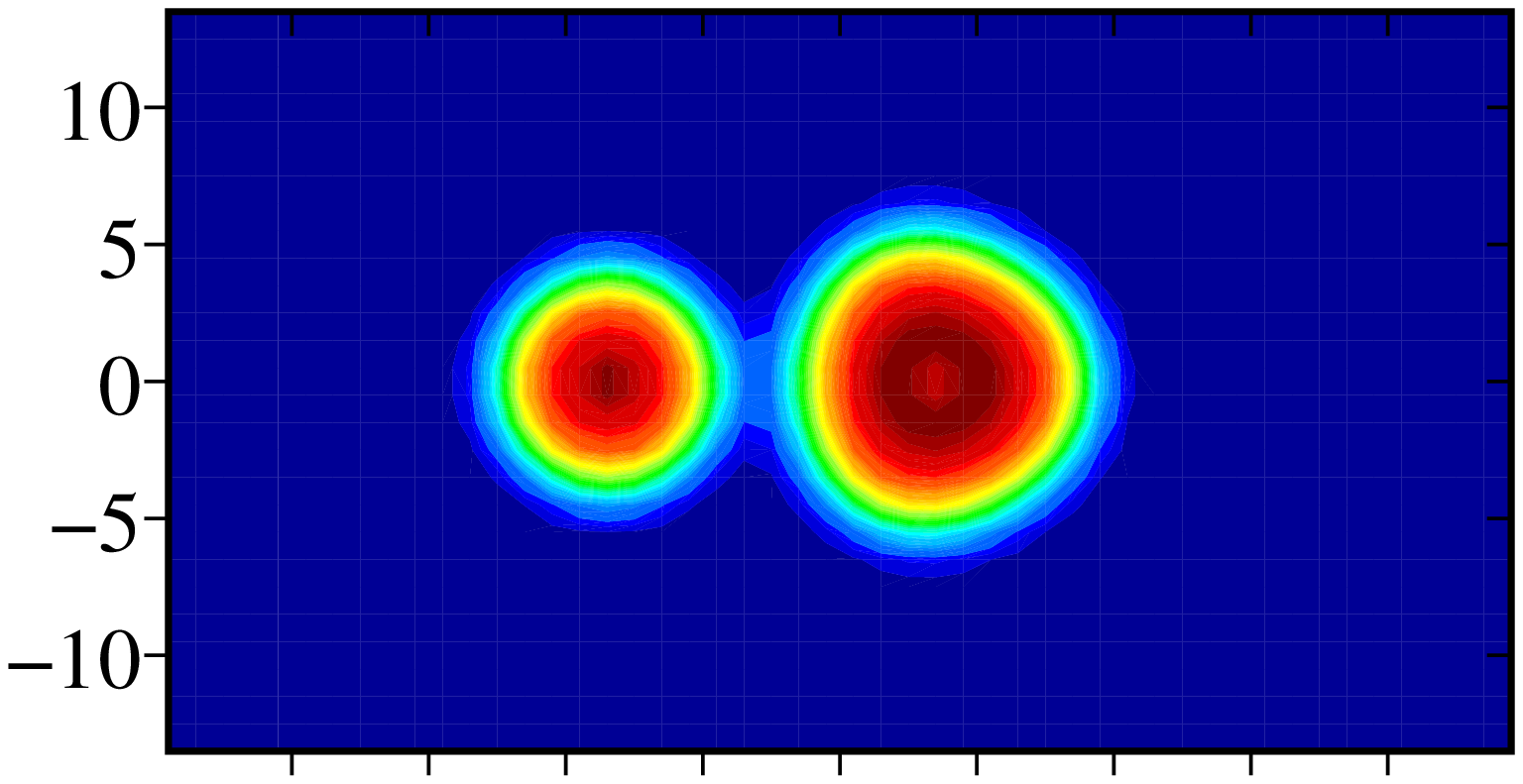}\vspace{-0.06in}\hspace{0.0in}\includegraphics*[scale=0.55]{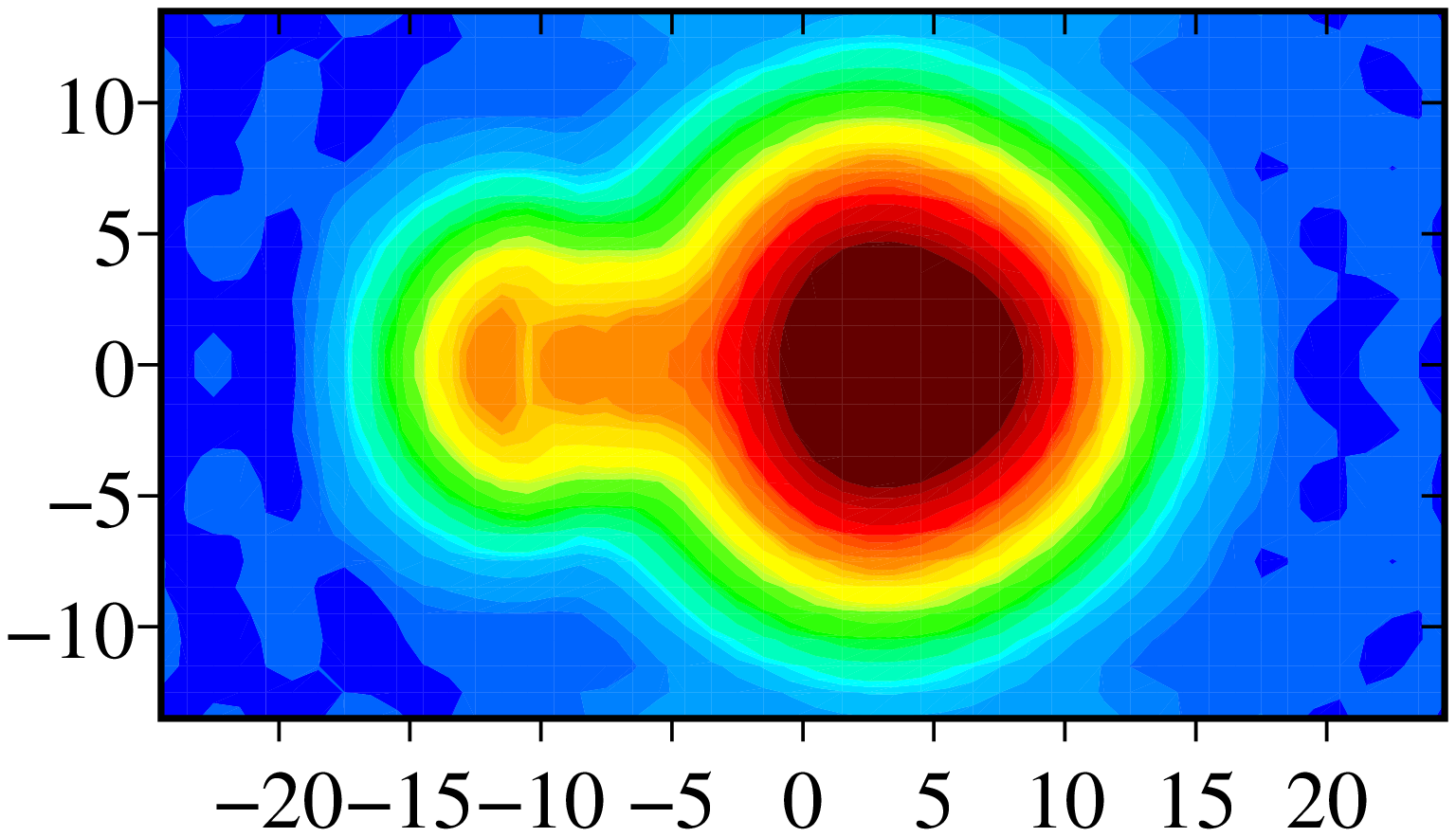}
\caption{\label{fig:dens} Contours of the total density for the dinuclear system
$^{40}$Ca+$^{96}$Zr at ion-ion separation of $R=12.3$~fm (top frame), and contours of the $^{96}$Zr
neutron states, plotted on a logarithmic scale at the same separation (bottom frame).}
\end{center}
\end{figure}

\begin{figure}[!htb]
\includegraphics*[scale=0.42]{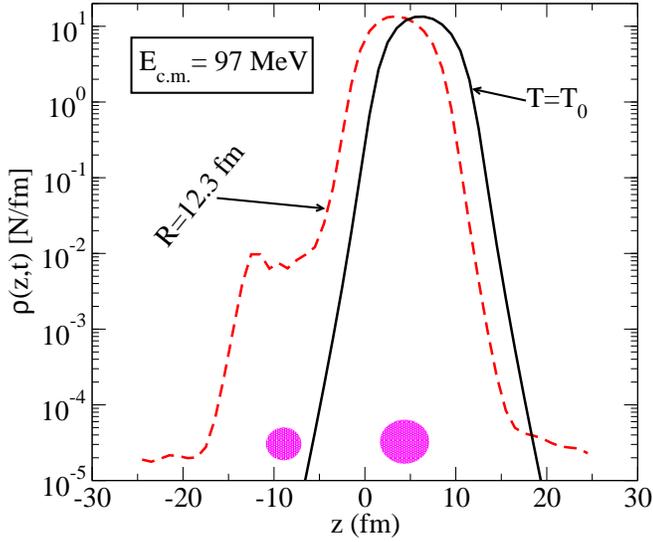}
\caption{\label{fig:rhoz2} Partially integrated neutron density calculated from Eq.(\protect\ref{eq:rhoz})
for the $^{96}$Zr nucleus
plotted on a logarithmic scale versus the collision axis coordinate $z$ for the $^{40}$Ca+$^{96}$Zr system
at $E_{\mathrm{c.m.}}=97$~MeV. The solid black curve corresponds to the initial partial
density and the dashed red curve to the same quantity at $R=12.3$~fm.}
\end{figure}

In Ref.~\cite{ZSW07} fusion for neutron-rich systems was studied in the vicinity of the Coulomb barrier 
using a semi-empirical time-dependent three-body Schr\"odinger equation. Their studies showed
that for the $^{40}$Ca+$^{96}$Zr system at $E_{\mathrm{c.m.}}=97$~MeV, which is in the vicinity
of the Coulomb barrier, neutrons were transferred in the early
stages of the collision ($R=$11-14~fm) from the $2d_{5/2}$ state of $^{96}$Zr to the unoccupied
levels of the $^{40}$Ca nucleus.
Here, we shall study the same system using the TDHF theory.
We have used the DC-TDHF method to find the effective potential barrier for this system as shown
in Fig.~\ref{fig:vrCaZr}. The barrier was calculated by initializing the TDHF run at $E_{\mathrm{c.m.}}=97$~MeV.
The barrier peak is about $95$~MeV and is located at about $R=11.5$~fm.
For this energy the TDHF collision results in fusion. In order to establish the early mass exchange we have
plotted the nuclear density and the corresponding neutron density of the $^{96}$Zr nucleus
(which is on the right half of the collision plane) in Fig.~\ref{fig:dens}. The neutron density was
plotted on a logarithmic scale to emphasize the low-density contours. The neutron tail seen in the
lower frame of Fig.~\ref{fig:dens} is very similar to the neutron density contours shown in Fig.~10 of
Ref.~\cite{ZSW07}.
For the same energy we have also calculated the partially integrated neutron density as a function of the
collision axis coordinate $z$ using Eq.(\protect\ref{eq:rhoz}) for the $^{96}$Zr nucleus, shown
in Fig.~\ref{fig:rhoz2}. In this case we do not have a recoiled final state since the system actually fuses.
The integrated neutron transfer at the ion-ion separation distance of about $R=12$~fm is approximately
$0.5$.

\begin{figure}[!htb]
\includegraphics*[scale=0.42]{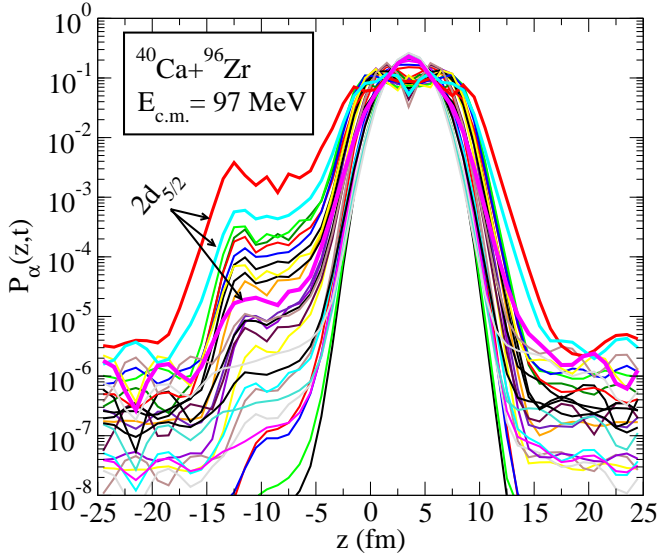}
\caption{\label{fig:palpha2} Neutron single-particle probability densities given by Eq.~(\protect\ref{eq:palpha})
for the $^{96}$Zr nucleus at $E_{\mathrm{c.m.}}=97$~MeV at $R=12.3$~fm.}
\end{figure}

To identify which states actually contribute to the probability of mass exchange from $^{96}$Zr to
$^{40}$Ca we have plotted the individual neutron single-particle probabilities given by Eq.~(\protect\ref{eq:palpha})
for the $^{96}$Zr nucleus as a function of collision coordinate axis $z$ at the same ion-ion separation
$R=12.3$~fm, as shown in Fig.~\ref{fig:palpha2}.
Again, the static Hartree-Fock calculation for the $^{96}$Zr nucleus helps us identify
these states. The states up to $N=50$ can be enumerated (by using parity, degeneracy of eigenvalues,
etc.) and exactly match the shell-model states with the spin-orbit term. The last six states (we do not
impose time-reversal invariance and thus have one state per nucleon) are almost degenerate in energy and
have positive parity values. This is consistent with Ref.~\cite{ZSW07}, which finds the neutron to be in
the $2d_{5/2}$ state. The corresponding three states (time-reversed pairs) are the red, cyan, and magenta
colored curves drawn thicker than others and pointed out by arrows. We observe that while two
of the states are the largest contributors to the transmission probability, one of the states practically
makes no contribution. This is true despite the fact that all these states are the highest energy
states and are degenerate in energy.
After the second highest $2d_{5/2}$ state at -7.05~MeV (cyan) the next three states
that have the largest contribution to the transmission probability
are; one of the two $2p_{3/2}$ states at -18.06~MeV, the $2p_{1/2}$ state at -16.17~MeV,
and only one of the five $1g_{9/2}$ states at -12.26~MeV.
The single-particle quadrupole moment is a good indicator of which of the states, specially sub-states
belonging to the same quantum number, make the largest contribution to the transmission probability,
since the most stretched states have the largest positive quadrupole moments.
For example the three substates of the $2d_{5/2}$ state have single-particle
quadrupole moments of $11.8$, $3.7$, and $-15.6$~fm$^2$ corresponding to the three
states shown in Fig.~\ref{fig:palpha2} in descending order. Similarly, the $2p_{3/2}$ state
with a quadrupole moment of $8.0$~fm$^2$ is the next largest contributor whereas the
other sub-state with a quadrupole moment of $-8.0$~fm$^2$ has a transmission probability
which is orders of magnitude smaller.
\begin{figure}[!htb]
\includegraphics*[scale=0.42]{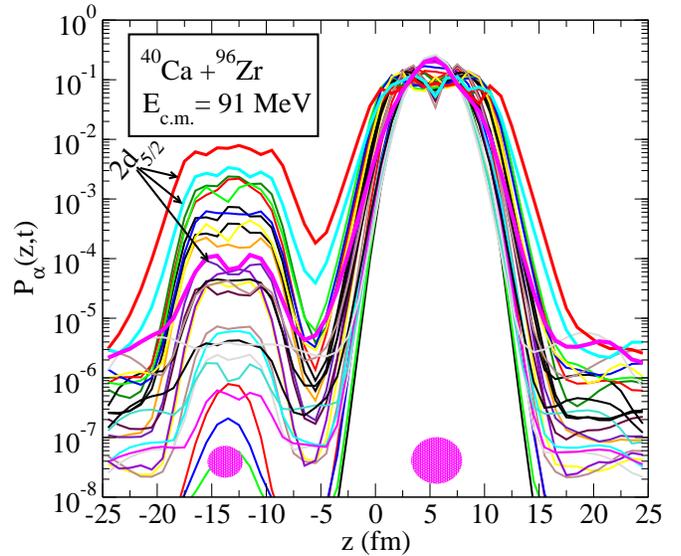}
\caption{\label{fig:palpha3} Neutron single-particle probability densities given by Eq.~(\protect\ref{eq:palpha})
for the $^{96}$Zr nucleus at $E_{\mathrm{c.m.}}=91$~MeV and after the recoiling nuclei are at a separation of
$R=19$~fm.}
\end{figure}
It is interesting to note that some of the sub-states of the same quantum number give
a much larger contribution, and states like $1f_{5/2}$, which is higher in energy
than $2p_{3/2}$, do not contribute appreciably.
This seems to indicate that the
transmission probability depends not only on the energies of the single-particle states, but
that it has an additional strong dependence on other properties of the states.
We have also repeated the same study for the  $^{40}$Ca+$^{96}$Zr system at
$E_{\mathrm{c.m.}}=91$~MeV, which is below the effective barrier. In this case
the nuclei recoil with a closest approach distance of $12.6$~fm.
Again, in order to identify which states actually contribute to the probability of mass exchange from $^{96}$Zr to
$^{40}$Ca we have plotted the individual neutron single-particle probabilities given by Eq.~(\protect\ref{eq:palpha})
for the $^{96}$Zr nucleus as a function of collision coordinate axis $z$ when the recoiled ions are about
$R=19$~fm apart, as shown in Fig.~\ref{fig:palpha3}.
We observe that the single-particle transmission probabilities
after the recoiled nuclei are well separated are similar to the
findings for the $E_{\mathrm{c.m.}}=97$~MeV case, except with reduced probabilities.
In addition, the observed sub-barrier behavior for the $^{40}$Ca+$^{96}$Zr system
is analogous to the $^{16}$O+$^{24}$O system.

\section{Conclusions}

We have performed a detailed analysis of mass exchange in the vicinity of the Coulomb
barrier for systems involving a neutron-rich nucleus using the TDHF theory. Our work
was motivated by Ref.~\cite{ZSW07} where the same phenomenon was studied
using a quantum mechanical three-body model. For the $^{40}$Ca+$^{96}$Zr system at
$E_{\mathrm{c.m.}}=97$~MeV, which is slightly above the barrier, we can essentially
confirm the results of Ref.~\cite{ZSW07}.
We confirm that at relatively large ion-ion distances the neutron transfer probability
begins to build up at the location of the receiving nucleus. We have analyzed this
aspect of neutron transmission in TDHF using the lighter $^{16}$O+$^{24}$O system.
We find that in the vicinity of the Coulomb barrier and below the barrier
different single-particle states see barriers that may differ from the effective
ion-ion barrier with a fixed center-of-mass. This reflects the fact
that the single-particle wave functions feel the barrier in the mean field
potential between the two nuclei, both concerning its height and its
position relative to their own geometric distribution, while the
ion-ion barrier is a bulk effect.
Due to this fact we find appreciable
transfer probability for energies well below the effective barrier.
This confirms that the incorporation of neutron transfer effects, as well
as other effects which depend on the properties of the single-particle states,
are necessary ingredients in fusion barrier calculations.

This work has been supported by the U.S. Department of Energy under grant No.
DE-FG02-96ER40963 with Vanderbilt University, and by the German BMBF
under contract No. 06 F 131.

\end{document}